\documentclass[11pt]{article}
\usepackage[margin=1in]{geometry}
\usepackage{amsmath,amssymb}
\usepackage{graphicx}
\usepackage{booktabs}
\usepackage{float}
\usepackage{natbib}
\usepackage{hyperref}
\usepackage{caption}
\usepackage{xcolor}
\usepackage{authblk}
\usepackage{times}
\setcitestyle{numbers,square,comma}
\hypersetup{colorlinks=true, linkcolor=blue, citecolor=blue, urlcolor=blue}

\title{\bfseries Resisting Drying Stress in Laponite--Bentonite Clay Composites: Two-Step Yielding and Crack--Peel Selection}

\author[1]{Tapati Dutta\thanks{Corresponding author.}}
\author[1]{Agnishikha Chakrabarti}
\author[1]{Soham Mitra}
\author[1]{Deepshikha Das}
\author[2]{Indranil Chakraborty}
\affil[1]{Physics Department, St.\ Xavier's College, Kolkata 700016, India}
\affil[2]{Chemistry Department, St.\ Xavier's College, Kolkata 700016, India}
\date{}

\begin{document}
\maketitle

\begin{abstract}
\noindent
Clay films that dry on a rigid substrate crack, peel or stay intact, and which of these happens matters for coatings, barrier layers and ceramic processing. Composites of clays with polymers have been studied widely; clay--clay composites, whose components differ in size and charge, are much less explored. We show that mixing the nanoscale clay Laponite with the micron-scale clay Bentonite reduces desiccation damage on both a hydrophobic and a hydrophilic substrate: Laponite films stay essentially intact, Bentonite films fracture extensively, and on glass most intermediate compositions are less damaged than a simple average of the two would suggest. Rheology of the gelled suspensions reveals a two-step yielding in every composition, and the cohesive energy derived from it depends by orders of magnitude on which step is used, a point that any rheology--cracking correlation must specify. A dimensionless cohesion--adhesion index, whose sign we derive from an energy balance, is consistent with compositions that peel on polypropylene and crack on glass. Replicate experiments were performed for robustness of results. The results identify composition as a simple handle on the crack resistance of clay films and open the search for the critical crack-free thickness as a function of clay composition.

\vspace{0.5em}
\end{abstract}

\section{Introduction}
\label{sec:intro}

Colloidal films that dry on a rigid substrate are put under tensile stress as they shrink, and relieve it by cracking, by peeling from the substrate, or not at all~\citep{routh2013,singh2007,man2008,goehring2013prl,goehring2015book}. Controlling which of these happens matters wherever clay layers are made by drying: protective and barrier coatings, engineered clay liners, ceramic green bodies and thin clay films in biomedical and sensing devices~\citep{bouazza2002,albrecht2001,gaharwar2014}. Much of the effort to improve the performance of clays has gone into composites of clays with polymers or organic molecules, in which the clay provides stiffness and barrier properties and the second component provides toughness~\citep{ray2003,pavlidou2008}. Composites of two clays, whose particles differ by an order of magnitude in size and in the distribution of charge, offer a different route, since both components are inorganic and both gel in water. To our knowledge their resistance to drying stress has not been examined systematically.

We study aqueous mixtures of two such clays. Laponite is a synthetic hectorite of disc-shaped nanoparticles with negatively charged faces and a pH-dependent positive rim, which gives attractive edge--face (``house-of-cards'') bonding and, at higher concentration or ionic strength, a more strongly arrested, electrostatically stabilised Wigner-glass-like state~\citep{ruzicka2011,jabbarifarouji2008}. Bentonite is a natural, montmorillonite-dominated clay whose particles are one to two orders of magnitude larger~\citep{luckham1999}. Earlier work from our group on single clays showed that cracking and peeling compete in drying clay films~\citep{sadhukhan2009,mal2007}, and that crack patterns depend on film thickness, with a critical thickness below which no cracks form~\citep{khatun2015}. Here we ask how the mixing ratio of two clays, and the substrate, set the outcome.

The paper makes four contributions. (i) Pure-Laponite and pure-Bentonite films, dried under the same conditions as the mixtures, are used as baselines against which the mixtures are compared (Section~\ref{sec:results-baselines}). (ii) Amplitude-sweep rheology of the as-gelled suspensions shows a two-step yielding in every composition, and we show that the cohesive energy scale derived from it changes by orders of magnitude with the step chosen (Section~\ref{sec:results-rheo}). (iii) Steady-shear viscosity curves of the gels show strong shear thinning with an apparent yield stress (Section~\ref{sec:results-flow}). (iv) A dimensionless cohesion--adhesion index, whose sign we derive from an energy balance, is tested against the observed failure modes (Sections~\ref{sec:theory-correction} and~\ref{sec:results-theory}). We use the rheology of the as-gelled suspension as a comparative probe of network cohesion and relate it to the drying outcome (Section~\ref{sec:discussion}).

\section{Materials and methods}
\label{sec:methods}

\subsection{Materials and film preparation}
\label{sec:methods-prep}
Natural Bentonite (analytical grade, purity $\geq99\%$, Sisco Research Laboratories) and Laponite RD (BYK Additives) were used. The Bentonite is montmorillonite-dominated (Al$_2$O$_3\cdot$4SiO$_2\cdot x$H$_2$O) with an average viscosity of $23{,}400\pm120.4$~mPa\,s and a shear-thinning index of 8.3519, that is, it is itself strongly shear thinning. Distilled water (conductivity $<1$~\textmu S\,cm$^{-1}$) was used throughout. The two clays were mixed at a fixed total clay concentration of 6.25~wt\% in the Laponite:Bentonite (L:B) mass ratios $r$ of Table~\ref{tab:compositions}; pure-Laponite and pure-Bentonite suspensions were prepared at the same concentration. Bentonite was pre-soaked in de-ionised water overnight, cooled to 20$^{\circ}$C, stirred for 5~min, combined with Laponite powder, and stirred for a further 30~s. The suspension was poured at once to a wet height of 1~mm into polypropylene (PP, inner diameter 5.3~cm) or glass (7.7~cm) petri dishes (Table~\ref{tab:dishes}) and dried at ambient conditions (RH $50\pm5\%$, $22\pm5^{\circ}$C) until no new cracks appeared over 24~h. The water contact angles of the substrates are $83.6^{\circ}$ (PP) and $29.4^{\circ}$ (glass). Crack patterns were photographed with a fixed camera setup (iPhone~7XR). Every composition was dried on each substrate in three sets of dishes under the same conditions; the images in Figure~\ref{fig:crackpatterns} are representative of one set and Figure~\ref{fig:replicates} shows the other two for five compositions. Damage scores (Section~\ref{sec:methods-score}) are means over the sets scored.

\begin{table}[htbp]
\centering
\caption{Dish geometry and film bookkeeping at a 1~mm wet pour height (suspension density 1.04~g/cm$^3$, 6.25~wt\% clay, clay density 2.6~g/cm$^3$). The dry thickness is the same in both dishes; the area and the perimeter-to-area ratio are not.}
\label{tab:dishes}
\small
\begin{tabular}{lcccccc}
\toprule
Substrate & Inner diameter (cm) & Area (cm$^2$) & Perimeter/area (cm$^{-1}$) & Suspension (mL) & Clay (g) & Dry $h$ (\textmu m) \\
\midrule
PP    & 5.3 & 22.1 & 0.755 & 2.21 & 0.143 & 25 \\
Glass & 7.7 & 46.6 & 0.519 & 4.66 & 0.303 & 25 \\
\bottomrule
\end{tabular}
\end{table}

Mass balance over the 6.25~wt\% solids loading and the clay density gives an idealized (fully dense) dry film thickness $h\approx25$~\textmu m in every dish, independent of dish diameter (Table~\ref{tab:dishes}). Real dried films retain some porosity, so this is a lower-bound estimate, used to make the cohesion--adhesion index dimensionless.

\begin{table}[htbp]
\centering
\caption{Compositions studied. $r=$~L:B mass ratio. Clay masses are for a 10~g batch (0.625~g total clay, 6.25~wt\% of suspension in every case; the masses of 1:5 and 1:4.5 are calculated from the ratio).}
\label{tab:compositions}
\small
\begin{tabular}{lcccc}
\toprule
L:B & $r$ & Laponite (g) & Bentonite (g) & Laponite (wt\% of susp.) \\
\midrule
Pure Bentonite & 0     & 0      & 0.625  & 0 \\
1:5   & 0.200 & 0.1042 & 0.5208 & 1.04 \\
1:4.5 & 0.222 & 0.1136 & 0.5114 & 1.14 \\
1:4   & 0.250 & 0.1250 & 0.5000 & 1.25 \\
1:3   & 0.333 & 0.1563 & 0.4688 & 1.56 \\
7:13  & 0.538 & 0.2188 & 0.4063 & 2.19 \\
1:1   & 1.000 & 0.3125 & 0.3125 & 3.13 \\
13:7  & 1.857 & 0.4063 & 0.2188 & 4.06 \\
3:1   & 3.000 & 0.4688 & 0.1563 & 4.69 \\
4:1   & 4.000 & 0.5000 & 0.1250 & 5.00 \\
4.5:1 & 4.500 & 0.5114 & 0.1136 & 5.11 \\
5:1   & 5.000 & 0.5208 & 0.1042 & 5.21 \\
Pure Laponite & $\infty$ & 0.625 & 0 & 6.25 \\
\bottomrule
\end{tabular}
\end{table}

\subsection{Rheology}
\label{sec:methods-rheo}
Rheology was performed \emph{in situ} at 22$^{\circ}$C on the freshly mixed suspension in an Anton Paar MCR~72 rheometer with a D-PP25 parallel-plate system, beginning once the sample had gelled, indicated by the suspension turning translucent between the plates, for every composition. This is a percolated gel, not a fluid sol; it is the pre-drying state at $\sim$2.5~vol\% solids, much more dilute than the consolidated packing present in the film when it cracks. Amplitude sweeps covered $\approx0.01$--$100\%$ strain. Steady-shear viscosity was recorded as time sweeps at a fixed shear rate of $50$~s$^{-1}$ (20 points over 120~s; L:B~=~1:3, 3:1, 1:5, 4:1) and as flow curves, that is, viscosity as a function of shear rate ($0.1$--$1000$~s$^{-1}$ for 1:1; $0.1$--$100$~s$^{-1}$ for 4:1 and 13:7). The 4:1 and 13:7 flow curves are reproduced from the instrument's graphical export: the reported viscosities at the first and last points and at 50~s$^{-1}$ are used exactly, and the intermediate points were read from the exported graph (reading the 50~s$^{-1}$ marker independently reproduced the reported values to within 0.2\%). SEM images were obtained for L:B~=~1:1, 3:1, 4:1 and 7:13.

\subsection{Damage scoring}
\label{sec:methods-score}
Desiccation cracking is sensitive to small changes in ambient conditions, so crack patterns are compared through a simple ordinal damage score for each dish: 0, intact; 1, rim-confined cracking only (or a hairline at the edge), interior intact; 2, edge cracking and/or detachment of the film from the wall over part of the perimeter, interior intact; 3, through-film interior cracking giving two to four fragments; 4, five or more fragments; 5, a dense polygonal network (ten or more cells). The dominant damage mode was recorded alongside. The score is a coarse tool for comparing patterns of different composition, not a precision measurement, and scores were assigned by one assessor from the photographs. The fragment-count thresholds are absolute and the glass dishes have 2.1 times the area of the PP dishes (Table~\ref{tab:dishes}), so score differences below about one unit between the two substrates are not interpreted.

\section{Rheological framework and cohesion--adhesion index}
\label{sec:theory}

These gels are viscoelastic: they respond partly like a solid and partly like a liquid, and how the mixture behaves under drying stress depends on how much energy the network can store, how readily it dissipates it, how it flows, and how strongly it is held by the substrate. The quantities introduced below are the standard ways of characterizing these aspects (the moduli and their crossovers, the viscosity, the conductivity and the work of adhesion), and the last subsection combines them into one dimensionless index.

\subsection{Moduli, crossovers and cohesive energy}
\label{sec:theory-rheo}

In oscillatory shear the storage modulus $G'$ measures the elastic response, the energy stored and returned on deformation, and the loss modulus $G''$ the dissipative response, the energy lost to irreversible rearrangement. Their ratio $\tan\delta=G''/G'$ says how solid-like ($<1$) or liquid-like ($>1$) the gel is. Within the linear regime the stored elastic energy density is $U_e\simeq\tfrac12G'\gamma^2$, and the strain $\gamma^{*}$ at which $G'=G''$ gives a comparative energy-storage scale $U_c=\tfrac12G'(0.1\%)\,\gamma^{*2}$.

Amplitude sweeps of this system do not, in general, cross from $G'>G''$ to $G'<G''$ only once. Processing the raw point-by-point record with one automated protocol, applied identically to every composition and detecting every sign change of $\log(G'/G'')$ over the measured strain range, shows that most compositions cross three times in sequence: an early crossover $\gamma_1^{*}$, an intermediate window in which $G'$ recovers above $G''$, and a terminal crossover $\gamma_2^{*}$ beyond which $G''>G'$ persists to the highest strain measured. Because $U_c$ depends on $\gamma^{*2}$, using $\gamma_1^{*}$ rather than $\gamma_2^{*}$ for the same composition can change $U_c$ by orders of magnitude (Section~\ref{sec:results-rheo}). We adopt the terminal crossover $\gamma_2^{*}$ as the single definition of $\gamma^{*}$ throughout, since it is the point beyond which the network does not re-elasticize within the measured window. We report the intermediate recovery through a recovery ratio $R=(G'/G'')_{\mathrm{peak}}/(G'/G'')_{\mathrm{min}}$ between $\gamma_1^{*}$ and $\gamma_2^{*}$.

\subsection{Viscous relaxation and electrical conductivity}
Steady-shear viscosity gives the characteristic relaxation time $\tau_r\sim\eta/G'$~\citep{coussot2005}; we use the viscosity at $50$~s$^{-1}$, the one shear rate at which every steady-shear measurement has a value, so $\tau_r$ is a comparative, not an absolute, relaxation time. Electrical conductivity $\sigma\propto\sum_iz_iu_ic_i$ probes the ionic environment~\citep{israelachvili2011,hunter2001}.

\subsection{Film--substrate adhesion}
The thermodynamic work of adhesion follows the Young--Dupr\'e relation $W_a=\gamma_L(1+\cos\theta)$. With $\gamma_L\approx72.8$~mN/m and the measured contact angles, $W_a(\mathrm{PP})\approx80.9$ and $W_a(\mathrm{glass})\approx136.2$~mJ/m$^2$~\citep{schrader1995,schrader1999}. This is the work of adhesion of water on the bare substrate, used as a wettability-based proxy for clay-film--substrate adhesion; a direct measurement (peel or blister testing of the dried film) would refine its magnitude but, as shown next, not the sign of its role in the index.

\subsection{A dimensionless cohesion--adhesion index}
\label{sec:theory-correction}

A ratio $U_c/(W_a\tan\delta)$, with $U_c$ in J/m$^3$ and $W_a$ in J/m$^2$, has units of inverse length; a length is needed to make it dimensionless, and the physically relevant one is the film thickness $h$:
\begin{equation}
\Psi = \frac{U_c\,h}{W_a\tan\delta}.
\label{eq:psi}
\end{equation}
The sign of the relation between $\Psi$ and failure mode follows from an energy balance. Cohesive cracking and interfacial delamination both relieve the same reservoir of drying-induced elastic energy; the mode selected is the one whose resistance is exceeded first. Delamination is resisted by the interfacial fracture energy, $\Gamma_i\approx W_a$: when $W_a$ is small the resistance is low and delamination is the easier path, favouring peeling; when $W_a$ is large, delamination is suppressed, stress stays in the film, and cohesive cracking becomes likely once the film's cohesive capacity is exceeded. Because $W_a$ is in the denominator of Eq.~\ref{eq:psi}, \textbf{large $\Psi$ (small $W_a$) favours peeling, and small $\Psi$ (large $W_a$) favours cohesive cracking.}

We test this sign against the data. For a given composition, $U_c$, $\tan\delta$ and $h$ are common to both substrates, so $\Psi$ differs between PP and glass only through $W_a$, and $\Psi(\mathrm{PP})/\Psi(\mathrm{glass})=1.68$ for every composition. A composition that fails by both mechanisms should therefore peel on PP and crack on glass, which Section~\ref{sec:results-theory} shows for L:B~=~1:1 and 13:7 in every dish examined. The two substrates were also dried in dishes of different diameter (Table~\ref{tab:dishes}), so this test cannot by itself separate adhesion from geometry (Section~\ref{sec:discussion}). $\Psi$ addresses the mode of failure, not its extent.

\section{Results}

\subsection{Crack patterns of the clay composites}
\label{sec:results-primary}
\begin{figure}[H]
\centering
\includegraphics[width=0.98\textwidth]{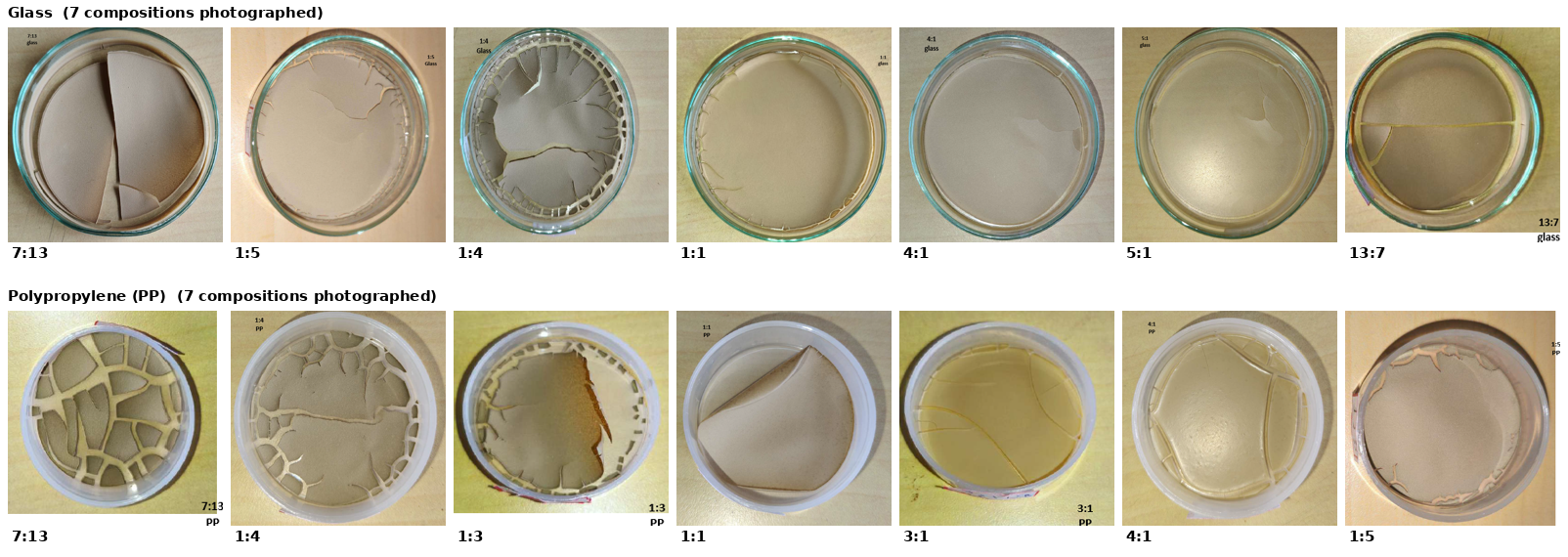}
\caption{Representative crack patterns (one of the three sets) on glass and PP for seven compositions. L:B mass ratio labeled per panel. Glass dishes are 7.7~cm and PP dishes 5.3~cm in inner diameter; panels are not drawn to a common scale.}
\label{fig:crackpatterns}
\end{figure}

Figure~\ref{fig:crackpatterns} shows representative crack patterns for seven compositions on each substrate, parameterized by $r=$~L:B. On PP, damage increases in the order 1:1 (peeling, no cracks) $<$ 4:1 (moderate peripheral ring) $\approx$ 3:1 (few long curved cracks) $<$ 1:5 (short edge cracks) $<$ 1:3 (jagged crack and rim ring) $<$ 1:4 (dense branched network) $<$ 7:13 (densest fragmentation). On glass, 4:1 has an intact interior with at most hairline rim cracking, 1:1 shows cracking confined to the pinned contact line at the rim, and 13:7 shows a near-linear crack.

Two features stand out across the compositions: films rich in Bentonite fragment into many pieces whereas films rich in Laponite stay largely intact, and films on PP tend to lift from the dish wall whereas films on glass remain pinned at the rim and crack there. The next section quantifies both.

\subsection{Pure clays, mixtures and reproducibility}
\label{sec:results-baselines}

\begin{figure}[htbp]
\centering
\includegraphics[width=0.98\textwidth]{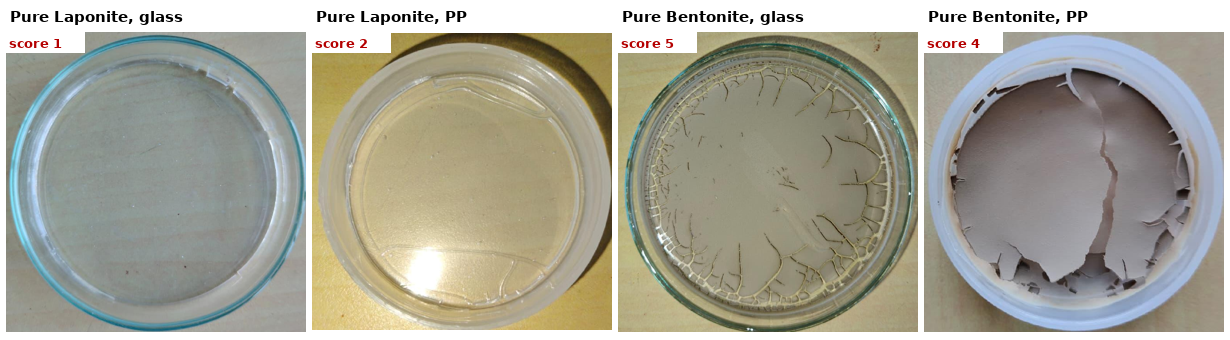}
\caption{Pure-component films (6.25~wt\%, 1~mm pour height) with damage scores.}
\label{fig:baselines}
\end{figure}

Figure~\ref{fig:baselines} shows the pure-component films. Pure Laponite forms a transparent film that stays essentially intact: on glass only short notches at the rim (score 1); on PP the film is pulled from the wall at the edge, with curled strips and one short crack, but the interior is intact (score 2). Pure Bentonite cracks extensively on both substrates: on glass a dense, edge-nucleated polygonal network with radial cracks running inwards (score 5); on PP the film splits through the centre and detaches from the wall into five or more fragments (score 4). Under identical loading, pour height and dishes the two clays therefore differ by three to four score units, with Laponite the crack-resistant component.

\begin{figure}[htbp]
\centering
\includegraphics[width=0.95\textwidth]{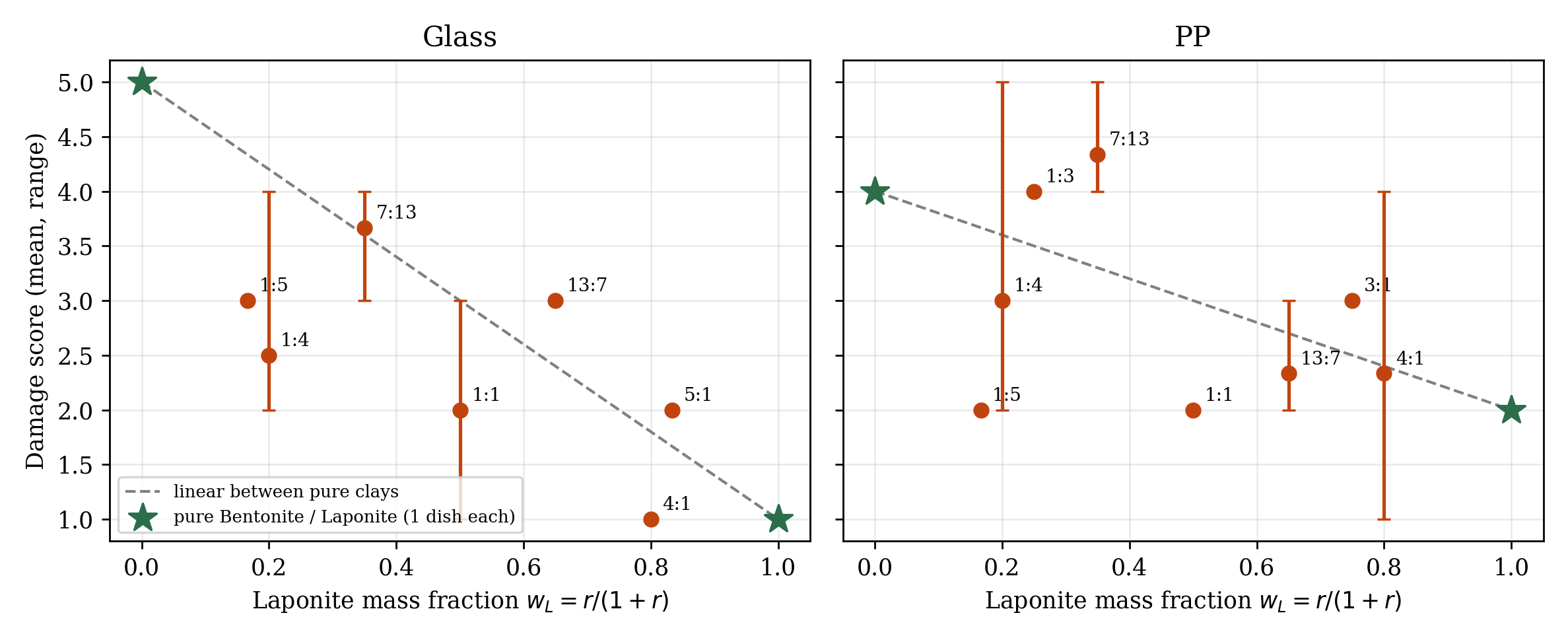}
\caption{Mean damage score (bars: range over the sets scored) versus Laponite mass fraction $w_L=r/(1+r)$ on glass and PP. Dashed line: linear interpolation between the pure-Bentonite and pure-Laponite scores.}
\label{fig:additivity}
\end{figure}

Figure~\ref{fig:additivity} compares the mixtures with a linear interpolation between the two pure-component scores, a reference for the absence of interaction, not a model. On glass, mixtures with 17--83~wt\% of the clay as Laponite are generally less damaged than the interpolation: the mean score of 1:5, 1:4, 1:1 and 4:1 lies 1.3, 1.7, 1.0 and 0.8 units below it, so that the first 20\% of Laponite lowers the score from 5 (pure Bentonite) to about 2.5 (1:4). Two compositions, 7:13 and 13:7, lie on or above the interpolation (+0.1 and +0.6), so the departure from additivity is not monotonic in composition. On PP there is no consistent departure (from $-1.7$ for 1:5 to $+1.0$ for 7:13). The mixtures are therefore not simply additive, particularly on the Bentonite-rich side on glass.

\begin{figure}[htbp]
\centering
\includegraphics[width=0.92\textwidth]{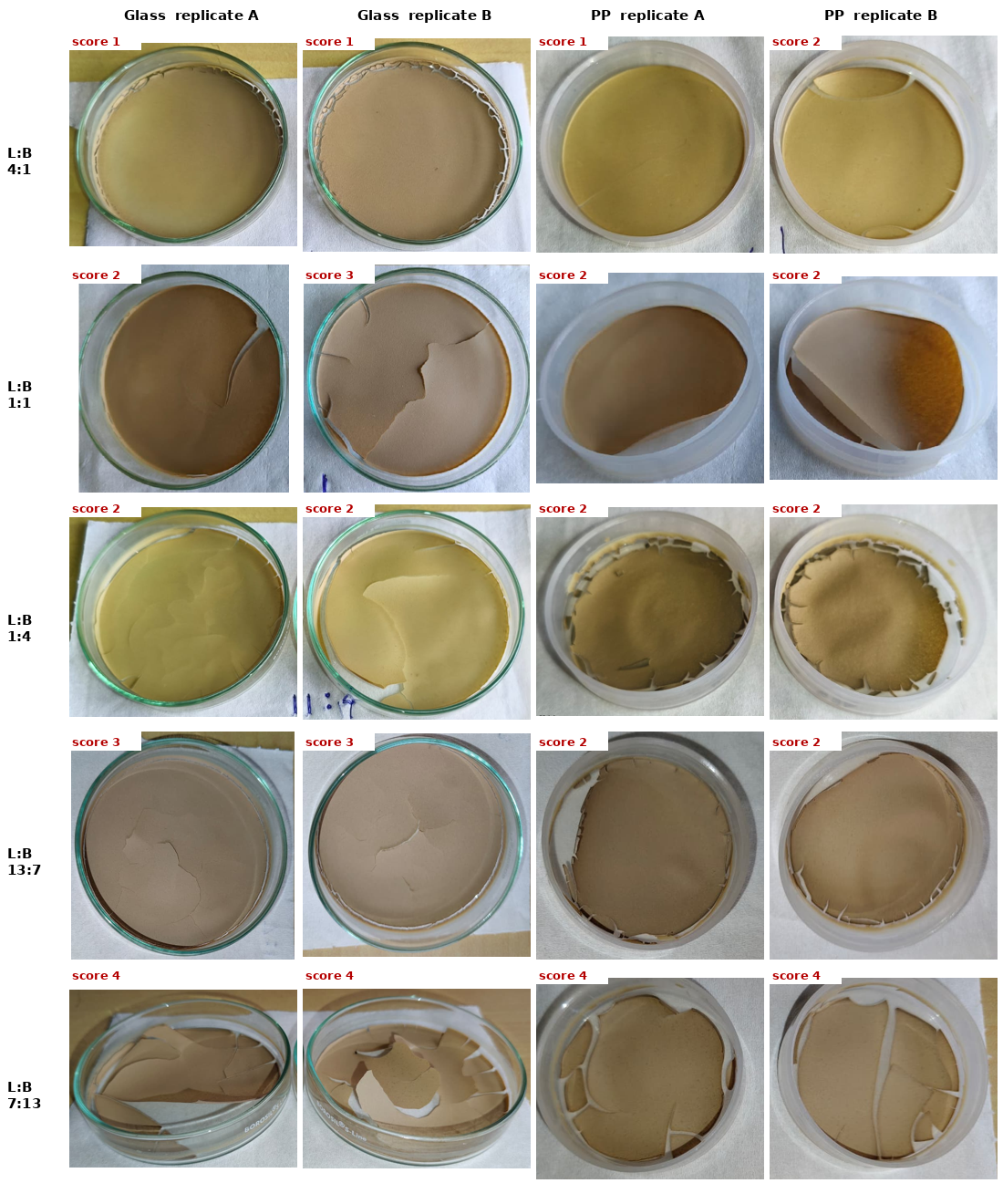}
\caption{The other two sets of dishes (A, B) for five compositions on both substrates, with damage scores. Compositions are ordered by increasing mean damage on glass. Panels are not drawn to a common scale.}
\label{fig:replicates}
\end{figure}

\begin{figure}[htbp]
\centering
\includegraphics[width=0.88\textwidth]{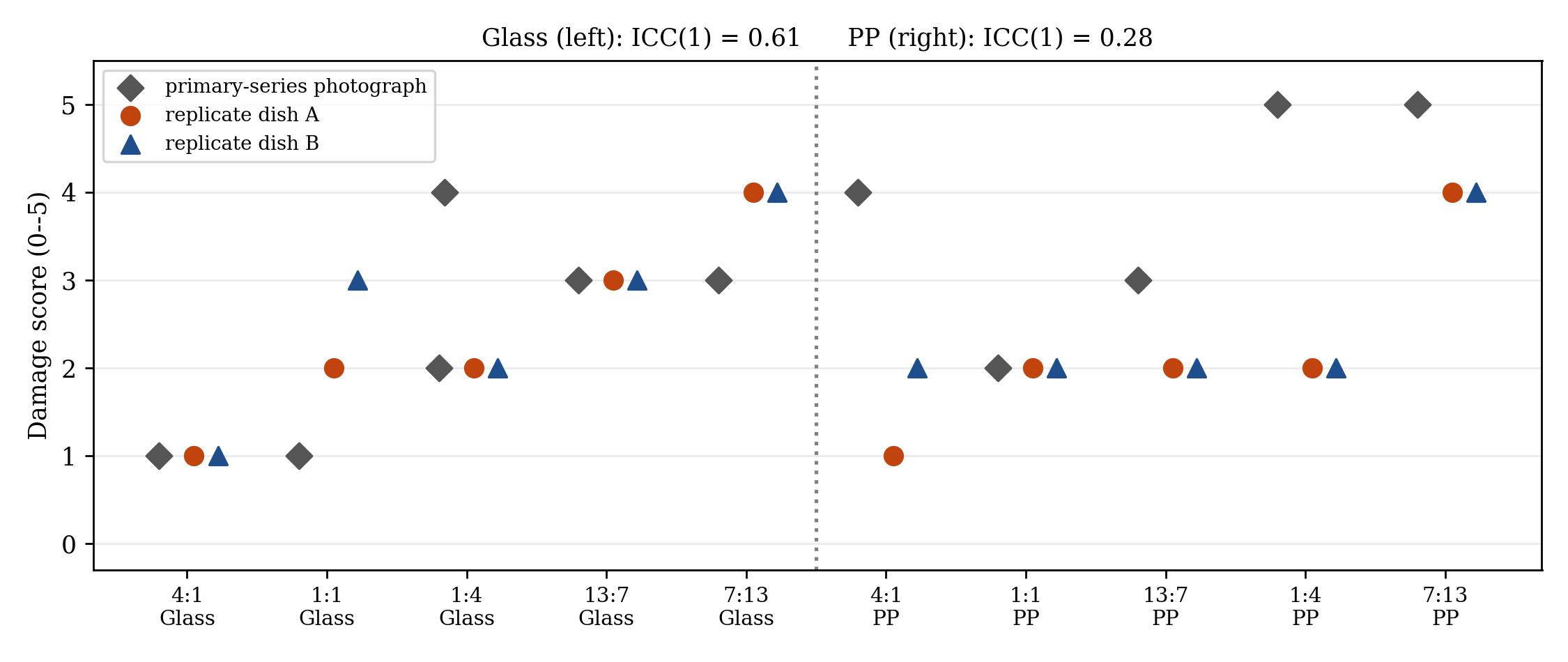}
\caption{Damage score of every dish for the ten conditions in Table~\ref{tab:repro} (primary-series photograph = set 1; dishes A and B = sets 2 and 3).}
\label{fig:scores}
\end{figure}

The three sets of dishes show the same main features and differ in their extent (Figures~\ref{fig:replicates} and~\ref{fig:scores}, Table~\ref{tab:repro}). Dishes dried side by side agree closely: eight of ten pairs have identical scores and the others differ by one unit. Across all three sets the intraclass correlation is 0.61 on glass and 0.28 on PP (0.41 overall; pooled within-condition standard deviation 0.92 score units). Three observations are reproduced in every set: 4:1 on glass is the least damaged condition, 7:13 is heavily fragmented on both substrates, and, on glass, the damage score falls steadily as the Laponite fraction rises (Spearman $\rho=-0.80$, $p=0.009$ over the nine compositions including both pure clays; PP $\rho=-0.38$, not significant). The largest difference between sets is on PP: the dense crack networks of 1:4 and 4:1 in the set of Figure~\ref{fig:crackpatterns} appear as films pulled from the wall, with few interior cracks, in the other two sets. Intermediate compositions on glass (1:1, 1:4, 13:7) differ by about one unit between sets and are not ordered by these data.

\begin{table}[htbp]
\centering
\caption{Damage scores of the three sets of dishes (set 1: Figure~\ref{fig:crackpatterns}; sets 2 and 3: Figure~\ref{fig:replicates}), with the damage mode in sets 2 and 3.}
\label{tab:repro}
\small
\begin{tabular}{llcccl}
\toprule
L:B & Substrate & Set 1 & Set 2 & Set 3 & Damage mode \\
\midrule
4:1  & Glass & 1    & 1 & 1 & rim cracking only; interior intact \\
1:1  & Glass & 1    & 2 & 3 & rim cracks; edge crack; one through-crack \\
1:4  & Glass & 2, 4 & 2 & 2 & partial detachment, rim cracking \\
13:7 & Glass & 3    & 3 & 3 & interior cracks, 3--5 fragments \\
7:13 & Glass & 3    & 4 & 4 & fragmentation, lifted pieces \\
\addlinespace
1:1  & PP    & 2    & 2 & 2 & film peeled or lifted, no cracks \\
4:1  & PP    & 4    & 1 & 2 & intact to edge lifting \\
13:7 & PP    & 3    & 2 & 2 & wall detachment, spiky rim cracks \\
1:4  & PP    & 5    & 2 & 2 & wall detachment, spiky rim cracks \\
7:13 & PP    & 5    & 4 & 4 & 4--6 fragments, wall gaps \\
\bottomrule
\end{tabular}
\end{table}

\subsection{Two-step yielding}
\label{sec:results-rheo}

\begin{figure}[htbp]
\centering
\includegraphics[width=0.98\textwidth]{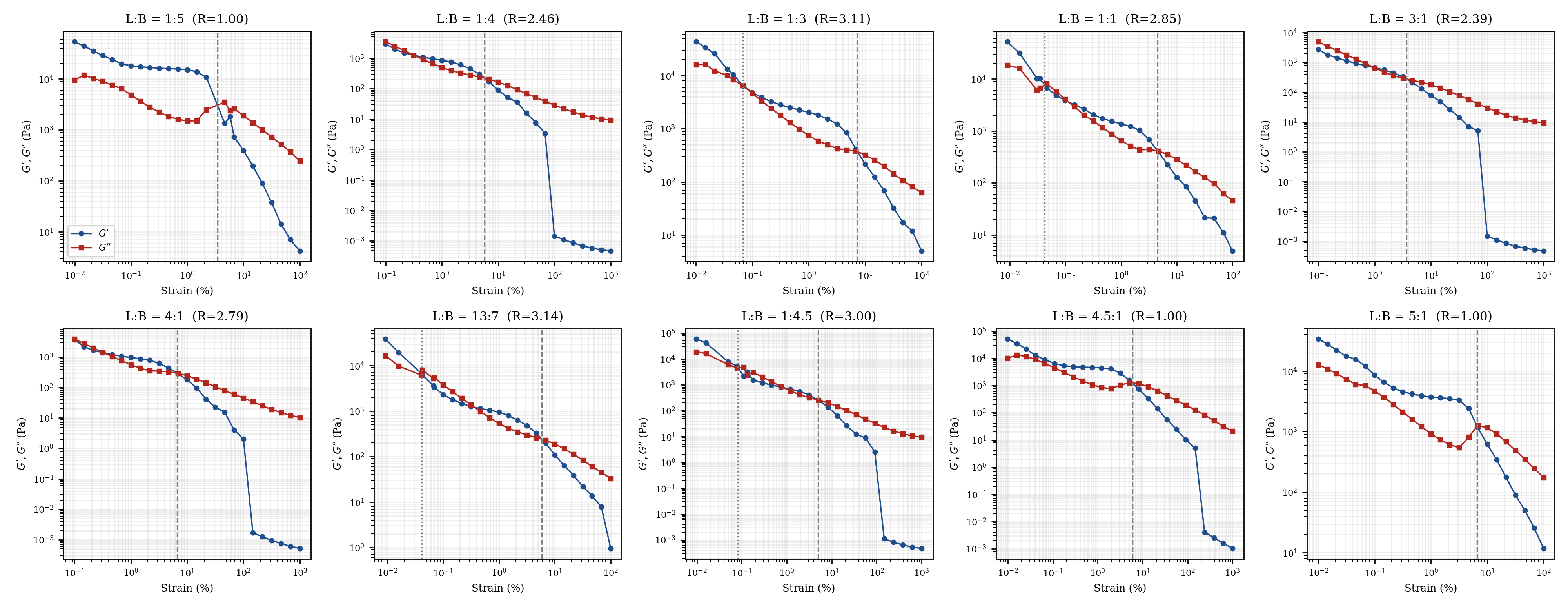}
\caption{Raw amplitude sweeps for the ten compositions with rheology. Dotted line: $\gamma_1^{*}$ (omitted where below the lowest strain measured); dashed line: $\gamma_2^{*}$, used for $U_c$ throughout. $R$ is given in each title.}
\label{fig:amplitude}
\end{figure}

Figure~\ref{fig:amplitude} shows the raw amplitude sweeps of all ten compositions in the instrument record, processed as described in Section~\ref{sec:theory-rheo}; Table~\ref{tab:rheo-sensitivity} lists the crossovers and the resulting $U_c$.

\begin{table}[htbp]
\centering
\caption{Crossover strains, recovery ratio $R$, and the dependence of $U_c$ on the crossover used. $G'$ and $\tan\delta$ at 0.1\% strain are used in $U_c=\tfrac12G'\gamma^{*2}$.}
\label{tab:rheo-sensitivity}
\small
\begin{tabular}{lcccccc}
\toprule
L:B & $\gamma_1^{*}$ (\%) & $\gamma_2^{*}$ (\%) & $R$ & $U_c(\gamma_1^{*})$ (J/m$^3$) & $U_c(\gamma_2^{*})$ (J/m$^3$) & $\tan\delta$ \\
\midrule
1:1   & 0.043 & 4.52 & 2.85 & $3.5\times10^{-4}$ & 3.96  & 1.04 \\
13:7  & 0.041 & 5.86 & 3.14 & $2.0\times10^{-4}$ & 3.99  & 1.64 \\
1:3   & 0.068 & 7.29 & 3.11 & $1.1\times10^{-3}$ & 12.85 & 0.97 \\
1:4.5 & 0.082 & 5.02 & 3.00 & $9.4\times10^{-4}$ & 3.49  & 1.71 \\
3:1   & $<$0.1 & 3.69 & 2.39 & -- & 1.84  & 1.81 \\
4:1   & $<$0.1 & 6.64 & 2.79 & -- & 8.09  & 1.06 \\
1:4   & $<$0.1 & 5.74 & 2.46 & -- & 4.91  & 1.17 \\
1:5   & 3.44  & 3.44 & 1.00 & = & 10.80 & 0.27 \\
4.5:1 & 5.89  & 5.89 & 1.00 & = & 11.75 & 0.71 \\
5:1   & 6.61  & 6.61 & 1.00 & = & 18.68 & 0.55 \\
\bottomrule
\end{tabular}

\vspace{0.3em}
\raggedright\footnotesize ``$<$0.1'': $\gamma_1^{*}$ lies below the lowest strain measured, so only $\gamma_2^{*}$ is resolved; ``='': a single crossover, so $\gamma_1^{*}=\gamma_2^{*}$.
\end{table}

\textbf{Every composition shows the same three-region structure}: an early yielding event, an intermediate window in which $G'$ recovers above $G''$, and a terminal crossover. The recovery ratio separates the ten compositions into two classes: $R\approx1$ (single-step: 1:5, 4.5:1, 5:1) and $R=2.4$--3.1 (double-step: 1:1, 1:3, 1:4, 1:4.5, 3:1, 4:1, 13:7). For four compositions (1:1, 13:7, 1:3, 1:4.5) both crossovers are resolved, and $U_c$ at $\gamma_1^{*}$ is $3.7\times10^{3}$ to $2\times10^{4}$ times smaller than at $\gamma_2^{*}$ for the same composition; for the others only one crossover is available. A per-composition choice of crossover would therefore spread $U_c$ over more than four decades, whereas with the terminal crossover applied to all ten it spans about one decade (1.8--18.7~J/m$^3$). The crossover must therefore be specified in any correlation of rheology with cracking, and the first crossover of a sweep cannot be taken as ``the'' crossover without checking whether a second exists.

We offer a hypothesis for the two steps, not a demonstrated mechanism. Bentonite platelets are roughly an order of magnitude larger than Laponite discs, and at every composition the number of Laponite particles, estimated from approximate particle volumes, far exceeds that of Bentonite even where Bentonite dominates by mass. A plausible reading is that $\gamma_1^{*}$ marks yielding of the numerous, low-energy Laponite edge--face bonds, while the recovery reflects a sparser, higher-energy-storage Bentonite-associated sub-structure that bears stress until $\gamma_2^{*}$.

\subsection{SEM microstructure}
\label{sec:results-sem}
\begin{figure}[htbp]
\centering
\includegraphics[width=0.75\textwidth]{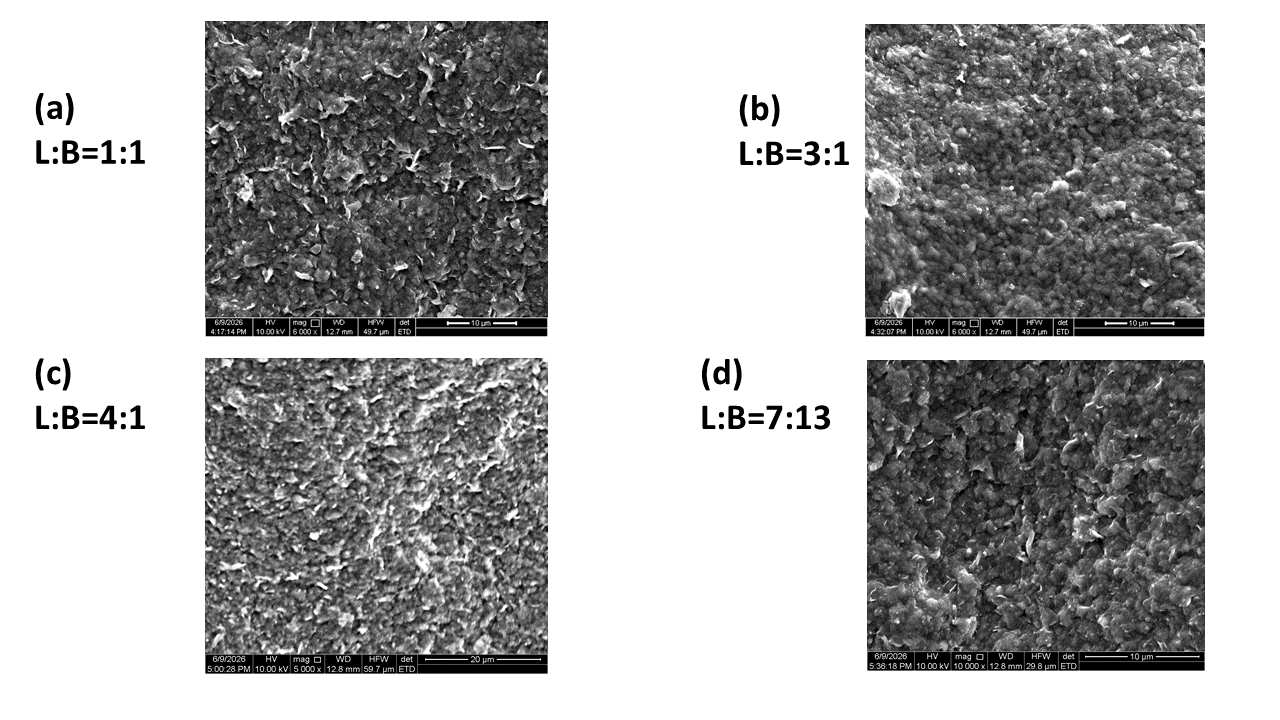}
\caption{SEM images of desiccated films, L:B~=~1:1, 3:1, 4:1, 7:13, at equal magnification (scale bar 10~\textmu m).}
\label{fig:sem}
\end{figure}

SEM images of the desiccated films at L:B~=~1:1, 3:1, 4:1 and 7:13 (Figure~\ref{fig:sem}), chosen as the compositions showing the least cracking in preliminary trials in order to look for a structural signature of crack resistance, show nano-sized Laponite discs arranged as stacked, cocci-shaped aggregates packed between the larger Bentonite particles, with no prominent grain dislocation or voids at the resolution used. At this magnification no feature distinguishes the four compositions from one another, although their crack outcomes differ substantially: 1:1, 3:1 and 4:1 look like 7:13, the most fragmented composition on both substrates. Whatever separates the drying behaviour of these compositions is therefore not a gross packing or void-fraction difference resolvable here, which points to the rate-dependent network response (Section~\ref{sec:results-rheo}) rather than to the static dried microstructure. Testing the two-step hypothesis requires a magnification-matched series with elemental mapping to distinguish Laponite- from Bentonite-rich regions.

\subsection{Flow curves and conductivity}
\label{sec:results-flow}

\begin{figure}[htbp]
\centering
\includegraphics[width=0.98\textwidth]{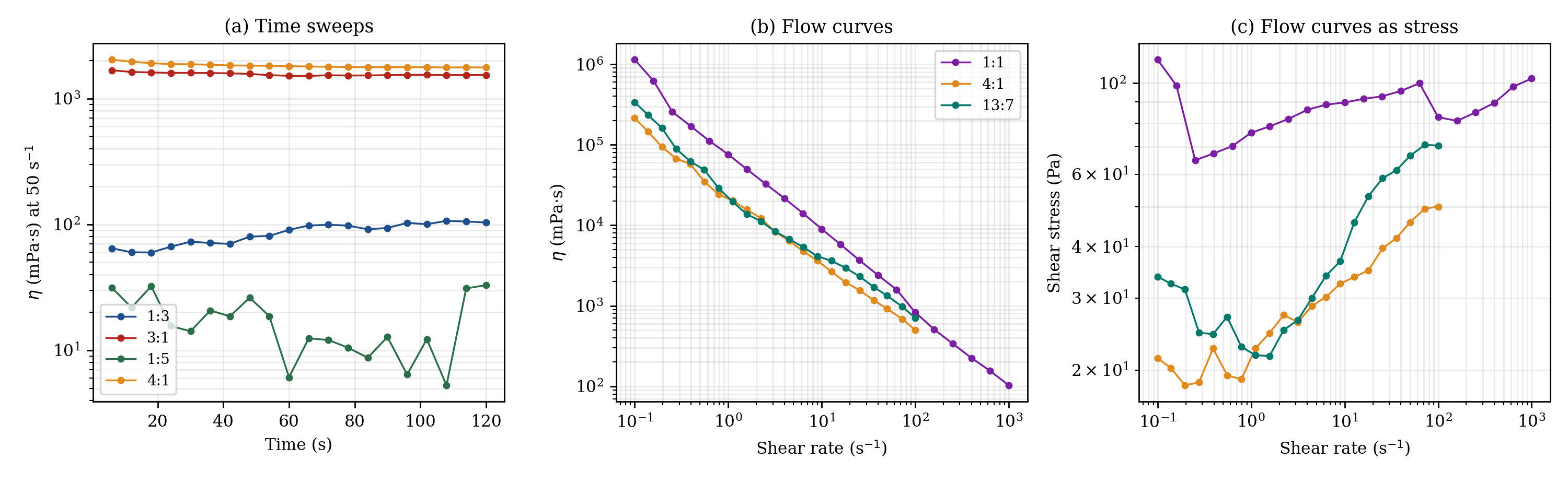}
\caption{(a) Viscosity at 50~s$^{-1}$ over 120~s for L:B~=~1:3, 3:1, 1:5 and 4:1. (b) Flow curves of 1:1, 4:1 and 13:7. (c) The same flow curves as shear stress.}
\label{fig:flow}
\end{figure}

Figure~\ref{fig:flow} and Table~\ref{tab:flow} give the steady-shear data: viscosity at a fixed shear rate of 50~s$^{-1}$ as a function of time (panel a) for four compositions, and viscosity as a function of shear rate, the flow curve (panels b, c), for three. All three flow curves show the same behaviour: the viscosity falls by 2.6--3.1 decades between 0.1 and 100~s$^{-1}$ while the shear stress changes little (1:1: 114 to 83~Pa; 4:1: 21 to 50~Pa; 13:7: 34 to 71~Pa), the signature of a gel with an apparent yield stress. Herschel--Bulkley fits ($\sigma=\sigma_y+K\dot\gamma^{n}$) give $\sigma_y\approx82$, 14 and 22~Pa for 1:1, 4:1 and 13:7, with $n=0.41$, 0.32 and 0.50 (root-mean-square log residual 0.14, 0.07, 0.17); we regard the low-rate stress as the more robust measure. The strong shear thinning is therefore shared by the three compositions measured and is of the kind shown by the Bentonite itself (shear-thinning index 8.3519, Section~\ref{sec:methods-prep}).

\begin{table}[htbp]
\centering
\caption{Viscosity at 50~s$^{-1}$, $G'$ at 0.1\% strain, and the comparative relaxation time $\tau_r=\eta(50\,\mathrm{s}^{-1})/G'$ for the compositions with steady-shear data. Time sweeps: mean $\pm$ SD of 20 points. Flow curves: interpolated at 50~s$^{-1}$.}
\label{tab:flow}
\small
\begin{tabular}{lccccc}
\toprule
L:B & Measurement & $\eta(50\,\mathrm{s}^{-1})$ (mPa\,s) & $G'$ (Pa) & $\tau_r$ (s) & Yielding class \\
\midrule
1:5  & time sweep & $17.5\pm9.2$ (52\% RSD) & 18230 & $9.6\times10^{-7}$ & single-step \\
1:3  & time sweep & $86\pm16$     & 4839 & $1.8\times10^{-5}$ & double-step \\
3:1  & time sweep & $1563\pm44$   & 2708 & $5.8\times10^{-4}$ & double-step \\
4:1  & time sweep & $1830\pm73$   & 3668 & $5.0\times10^{-4}$ & double-step \\
4:1  & flow curve & 917           & 3668 & $2.5\times10^{-4}$ & double-step \\
1:1  & flow curve & 1956          & 3865 & $5.1\times10^{-4}$ & double-step \\
13:7 & flow curve & 1331          & 2318 & $5.7\times10^{-4}$ & double-step \\
\bottomrule
\end{tabular}
\end{table}

The double-step compositions with a measured viscosity, other than 1:3, have $\tau_r$ of 2.5--5.8~$\times10^{-4}$~s, about 260 to 600 times longer than the single-step composition 1:5; 1:3, also double-step, sits between ($1.8\times10^{-5}$~s). This is a second rheological signature consistent with the two-step picture, not a clean split on $\tau_r$ alone. For 4:1 the flow curve gives a viscosity at 50~s$^{-1}$ half that of the time sweep (917 against 1830~mPa\,s): the flow curve reaches 50~s$^{-1}$ after a ramp through lower rates and the time sweep from rest, which shows how sensitive the viscosity is to shear history. The viscosity of 1:3 rises smoothly during the sweep whereas that of 1:5 fluctuates by 52\%, which we read as intermittent restructuring of the more Laponite-rich, higher-$G'$ network of 1:5 under continued shear.

Mean electrical conductivity (relative units) decreases with increasing Laponite content over the four compositions measured, $\approx3.15$ (1:3), 2.35 (1:1), 2.35 (4:1), 1.65 (3:1), consistent with progressively more counterions being associated with the clay network. The accuracy of the conductivity meter was of the order of the run-to-run fluctuation of the recorded traces, so only the mean is used.

\subsection{The index against observed outcomes}
\label{sec:results-theory}

Because $U_c$, $\tan\delta$ and $h$ are common to both substrates, $\Psi(\mathrm{PP})/\Psi(\mathrm{glass})=1.68$ for every composition, and the sign argument of Section~\ref{sec:theory-correction} predicts peeling on PP and cracking on glass for a composition that fails by both. This is what is observed for 1:1: on PP the film peeled or lifted without cracking in all three sets (scores 2, 2, 2), whereas on glass it cracked in all three (rim cracks, an edge crack penetrating about a third of the diameter, and a through-crack; scores 1, 2, 3). It is also what 13:7 shows: on glass, interior cracking giving three to five fragments in all three sets; on PP, detachment from the wall with spiky edge cracks and an intact interior in the two further sets.

\begin{figure}[htbp]
\centering
\includegraphics[width=0.78\textwidth]{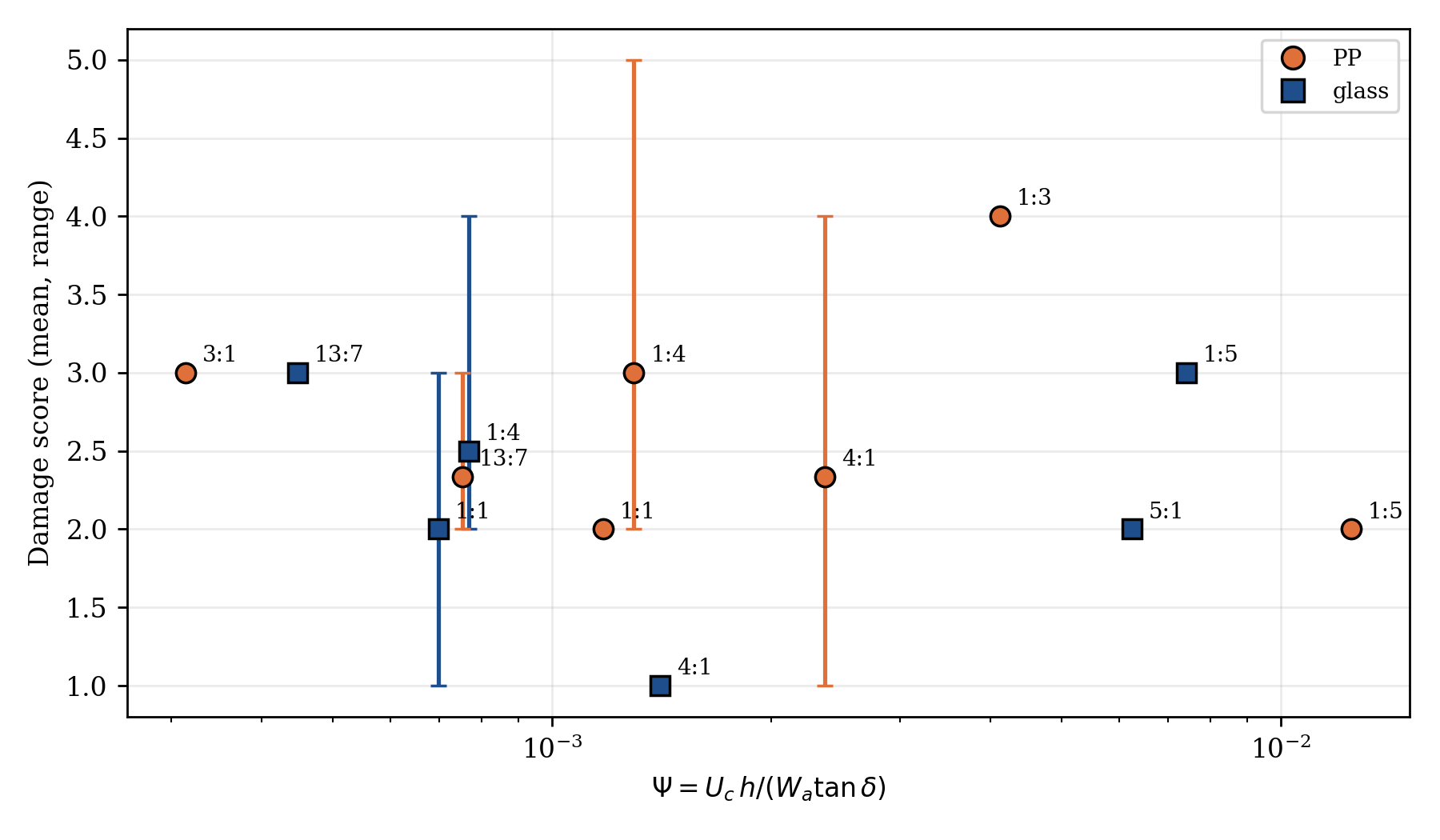}
\caption{Mean damage score (bars: range over the sets scored) versus $\Psi=U_ch/(W_a\tan\delta)$ for the 13 composition--substrate pairs with rheology and photographs ($h=25$~\textmu m, $U_c$ at $\gamma_2^{*}$). Within each composition the PP point lies at $\Psi$ 1.68 times that of the glass point.}
\label{fig:psicorr}
\end{figure}

Figure~\ref{fig:psicorr} plots the mean damage score against $\Psi$ for the 13 composition--substrate pairs with rheology and photographs. The two are uncorrelated (Spearman $\rho=-0.16$, $p=0.61$), as expected: $\Psi$ addresses the mode of failure (peel against crack), whereas the damage score measures extent, which also depends on film thickness, the flaw population and the boundary conditions at the pinned contact line.

\section{Discussion}
\label{sec:discussion}

\subsection{A clear signal from a comparative probe}
The rheology was measured on a gelled suspension at $\sim$2.5~vol\% solids, at shear rates and strains far above those of drying: for a $\sim$1~mm wet film evaporating at $\sim$1~mm/day the strain rate is $(1/h)(dh/dt)\approx8\times10^{-6}$~s$^{-1}$, rising to $\approx2\times10^{-4}$~s$^{-1}$ as the film thins, and the film at fracture is a consolidated packing at tens of vol\% governed also by capillary pressure, particle size and thickness~\citep{singh2007,man2008}. The absolute values of $\gamma^{*}$ and $U_c$ are therefore not predictions of the stress at fracture. A clear signal nevertheless emerges, in which the characteristics of the two clays and those of the substrate combine. First, damage is controlled by composition: on glass it falls steadily with the Laponite fraction ($\rho=-0.80$), and the two pure clays differ by three to four score units, so the particles themselves, with their different size and charge, set how much stress the film can bear. Second, the substrate sets the mode: within a composition that fails by both routes, the film peels where adhesion is weak and cracks where it is strong, as the sign of $\Psi$ requires. Third, the rheology identifies the structure common to all the gels, a two-step yielding, and the cohesive energy $U_c$ derived from it enters $\Psi$ together with the adhesion $W_a$ and the dissipation $\tan\delta$. $U_c$ alone does not order the extent of damage (Spearman $|\rho|<0.2$ on either substrate, over a tenfold range of $U_c$), and need not: it is the balance of cohesion (clay composition), adhesion (substrate) and dissipation that decides whether stress is relieved by cracking or peeling. That a comparative ranking is informative despite the differences in rate and concentration is consistent with the gel network being the precursor of the consolidated film, so that its relative cohesion is inherited; thickness- and rate-resolved measurements can test this.

\subsection{Substrate, dish size and the scatter of cracking}
Substrate and dish diameter differ together in this work (PP: 5.3~cm, perimeter-to-area ratio 0.755~cm$^{-1}$; glass: 7.7~cm, 0.519~cm$^{-1}$; Table~\ref{tab:dishes}). The glass dishes have 2.1 times the area, which raises fragment counts at equal crack spacing, and the PP dishes 1.45 times the perimeter-to-area ratio, which favours edge-initiated detachment, so peel-against-crack contrasts cannot be attributed to adhesion alone. The cleanest remedy is to remove the difference, by using dishes of the same diameter on both substrates, for example a PP film in a glass dish of the same diameter. Two alternatives allow the existing data to be normalized. Because crack spacing in a drying film scales with its thickness, the number of fragments should scale with the area measured in units of $h^2$, so fragment counts can be divided by dish area, and edge damage by perimeter; for 7:13 the approximate interior fragment density is then 0.23--0.27~cm$^{-2}$ on PP against 0.13--0.15~cm$^{-2}$ on glass, whereas for 13:7 it is 0.09--0.11~cm$^{-2}$ on glass against 0.05~cm$^{-2}$ on PP, so the normalization does not give a consistent substrate difference and the sign of the contrast depends on composition. A diameter series on one substrate would measure the diameter dependence directly and allow results to be referred to a common diameter. Desiccation cracks are sensitive to small changes in temperature, humidity and handling, which is why three sets of dishes were used; with the scatter we observe (pooled standard deviation 0.92 score units), differences of about two score units are resolved by three sets, and a difference of one unit would need about thirteen. The composition trends described here should be read at that level.

\section{Conclusions}
This work shows that composition is a simple handle on the resistance of clay films to drying stress. Mixing the nanoscale clay Laponite with the micron-scale clay Bentonite gives films that are mostly less damaged than a simple average of the two pure clays would suggest, particularly on glass, with Laponite the crack-resistant component and Bentonite the crack-prone one. The amplitude-sweep rheology of every composition shows a two-step yielding, and the cohesive energy derived from it depends by up to four decades on which step is used; fixing the terminal step gives cohesive energies within about one decade. The strong shear thinning with an apparent yield stress of the gels is shared by the compositions measured. A dimensionless cohesion--adhesion index, whose sign follows from an energy balance, is consistent with compositions that peel on polypropylene and crack on glass.

These results are relevant wherever clay layers are made by drying: protective and barrier coatings, clay liners that must resist desiccation, and clay films for biomedical and sensing devices. Because both components are inorganic and gel in water, tuning the mixing ratio offers a route to crack-resistant clay films that does not require a polymer or organic additive.

The next step is the critical crack-free thickness as a function of clay composition, which we are investigating and will report separately; it will show how much of the crack resistance reported here is intrinsic to the composition. Dishes of equal diameter on both substrates, characterization of the Bentonite (cation-exchange capacity and exchangeable cations), and rheology at drying-relevant rates will connect the comparative picture here to a predictive one.

\section*{Acknowledgement}
AC is grateful to the SEM facility of the Physics Department at Jadavpur University.

\section*{Data availability}
Raw rheometer output files and the full set of primary crack photographs are available from the corresponding author upon reasonable request.


\begin{thebibliography}{99}
\bibitem{routh2013} A. F. Routh, \textit{Rep. Prog. Phys.}, 2013, \textbf{76}, 046603.
\bibitem{singh2007} K. B. Singh and M. S. Tirumkudulu, \textit{Phys. Rev. Lett.}, 2007, \textbf{98}, 218302.
\bibitem{man2008} W. Man and W. B. Russel, \textit{Phys. Rev. Lett.}, 2008, \textbf{100}, 198302.
\bibitem{goehring2013prl} L. Goehring, W. J. Clegg and A. F. Routh, \textit{Phys. Rev. Lett.}, 2013, \textbf{110}, 024301.
\bibitem{goehring2015book} L. Goehring, A. Nakahara, T. Dutta, S. Kitsunezaki and S. Tarafdar, \textit{Desiccation Cracks and their Patterns: Formation and Modelling in Science and Nature}, Wiley-VCH, 2015.
\bibitem{bouazza2002} A. Bouazza, \textit{Geotext. Geomembr.}, 2002, \textbf{20}, 3--17.
\bibitem{albrecht2001} B. A. Albrecht and C. H. Benson, \textit{J. Geotech. Geoenviron. Eng.}, 2001, \textbf{127}, 67--75.
\bibitem{gaharwar2014} A. K. Gaharwar, N. A. Peppas and A. Khademhosseini, \textit{Biotechnol. Bioeng.}, 2014, \textbf{111}, 441--453.
\bibitem{ray2003} S. Sinha Ray and M. Okamoto, \textit{Prog. Polym. Sci.}, 2003, \textbf{28}, 1539--1641.
\bibitem{pavlidou2008} S. Pavlidou and C. D. Papaspyrides, \textit{Prog. Polym. Sci.}, 2008, \textbf{33}, 1119--1198.
\bibitem{ruzicka2011} B. Ruzicka and E. Zaccarelli, \textit{Soft Matter}, 2011, \textbf{7}, 1268--1286.
\bibitem{jabbarifarouji2008} S. Jabbari-Farouji, H. Tanaka, G. H. Wegdam and D. Bonn, \textit{Phys. Rev. E}, 2008, \textbf{78}, 061405.
\bibitem{luckham1999} P. F. Luckham and S. Rossi, \textit{Adv. Colloid Interface Sci.}, 1999, \textbf{82}, 43--92.
\bibitem{sadhukhan2009} S. Sadhukhan, J. Prehl, P. Blaudeck, K. H. Hoffmann, T. Dutta and S. Tarafdar, \textit{Eur. Phys. J. E}, 2009, \textbf{27}, 391--395.
\bibitem{mal2007} D. Mal, S. Sinha, T. Dutta, S. Mitra and S. Tarafdar, \textit{J. Phys. Soc. Jpn.}, 2007, \textbf{76}, 014801.
\bibitem{khatun2015} T. Khatun, T. Dutta and S. Tarafdar, \textit{Eur. Phys. J. E}, 2015, \textbf{38}, 83.
\bibitem{coussot2005} P. Coussot, \textit{Rheometry of Pastes, Suspensions, and Granular Materials}, Wiley, New York, 2005.
\bibitem{israelachvili2011} J. N. Israelachvili, \textit{Intermolecular and Surface Forces}, Academic Press, Oxford, 3rd edn, 2011.
\bibitem{hunter2001} R. J. Hunter, \textit{Foundations of Colloid Science}, Oxford University Press, Oxford, 2nd edn, 2001.
\bibitem{schrader1995} M. E. Schrader, \textit{Langmuir}, 1995, \textbf{11}, 3585--3589.
\bibitem{schrader1999} M. E. Schrader, \textit{J. Colloid Interface Sci.}, 1999, \textbf{213}, 602--605.
\end{thebibliography}
\end{document}